# Why do Collapsed Carbon Nanotubes Twist?

Hamid Reza Barzegar[1,2,3,†], Aiming Yan[1,3,†], Sinisa Coh[1,3], Eduardo Gracia-Espino[2], Claudia Ojeda-Aristizabal[1,3,§], Gabriel Dunn[1,3], Marvin L. Cohen[1,3], Steven G. Louie[1,3], Thomas Wågberg[2] and Alex Zettl[1,3,4,*]

[1]Department of Physics, University of California, Berkeley, CA 94720, USA
[2]Department of Physics, Umea University, 90187 Umea, Sweden
[3]Materials Sciences Division, Lawrence Berkeley National Laboratory, Berkeley, CA 94720, USA
[4]Kavli Energy NanoSciences Institute at the University of California, Berkeley and the Lawrence Berkeley National Laboratory, Berkeley, CA 94720, USA

[†] Both authors contributed equally to this work
[§] Present Address: Department of Physics & Astronomy, California, State University Long Beach, Long Beach, California 90840, USA
[*] Address correspondence to azettl@berkeley.edu

ABSTRACT:

We study the collapsing and subsequent spontaneous twisting of a carbon nanotube by *in-situ* transmission electron microscopy. A custom-sized nanotube is first created in the microscope by selectively extracting shells from a parent multi-wall tube. The few-wall, large-diameter daughter nanotube is driven to collapse via mechanical stimulation, after which the ribbon-like collapsed tube spontaneously twists along its long axis. *In-situ* diffraction experiments fully characterize the uncollapsed and collapsed tubes. From the experimental observations and associated theoretical analysis, the origin of the twisting is determined to be compressive strain due to charge imbalance.

Collapsed carbon nanotubes (CCNTs)[1] share attributes of conventional (i.e. inflated or cylindrical) nanotubes from which they are typically derived, as well as attributes of graphene nanoribbons which they closely resemble geometrically. Indeed, a collapsed nanotube may be viewed as a realization of the atomically-perfect multi-layer graphene nanoribbon, in that all edge atoms are fully bonded without hydrogenation or other functionalization. CCNTs have been studied extensively experimentally and theoretically[2-11]. The physical origin of CCNTs is straightforward[1]: a conventional carbon nanotube minimizes its elastic curvature energy by assuming a cylindrical shape, but, if the nanotube has a large enough diameter and few enough shells (i.e. walls), the van der Waals energy tapped by allowing opposing "faces" of the inner walls to come into close proximity and sticking together can overwhelm the increase in strain energy at the edges, leading to a metastable or even globally stable collapsed state. A collapsed single-wall carbon nanotube resembles a bilayer graphene nanoribbon, while a collapsed double-wall carbon nanotube resembles a 4-layer graphene nanoribbon, etc. Importantly, the chirality of the innermost shell of the uncollapsed nanotube puts severe constraints on the allowed layer-to-layer stacking of adjacent inner layers in the collapsed tube.

Similar to the case of graphene nanoribbons, CCNTs can undergo additional structural changes such as curling, folding, or, rather strikingly, a twisting along their long axis with a well defined periodicity[12,13]. Although several theories have been advanced to account for the twisting in CCNTs[12,14,15] as well as in graphene nanoribbons[16,17], there is no consensus on the dominant driving mechanism. The experimental situation is even less clear. CCNTs are often observed accidentally, with little understanding of the dynamics of the formation/twisting process or the sensitivity of those processes to nanotube geometry. No real-time transmission electron microscopy (TEM) tracking of such processes has been reported.

Here we employ *in-situ* TEM to track the collapse and subsequent spontaneous twisting of a carbon nanotube in vacuum. A nanomanipulator is first used to create a custom-sized nanotube in the microscope, by telescopically extracting shells from a multi-wall parent nanotube. The daughter nanotube has suitable diameter and wall number so that it is prone to collapse, and it is induced to do so via mechanical perturbation from the nanomanipulator. The CCNT is then singly clamped at its end, and, while suspended in vacuum, it is observed to spontaneously twist along its long axis. By means of TEM imaging and electron diffraction we track the structural changes during collapsing and twisting. We use the experimental observations to test various theoretical models for twisting, and show that the main driving force for twisting of a CCNT is compressive strain.

The left panel drawings in Fig. 1 show schematically our experimental process to create a CCNT *in-situ* by nano-manipulation. Initially, one end of the parent conventional inflated multi-wall carbon nanotube (right side in Fig. 1(a)) is fixed via silver paint to a copper mesh TEM grid, which is then attached to a stationary sample holder in the TEM. Next, the opposite (left) end of the parent nanotube is spot-welded *in-situ* via a bias voltage (typically between 1 and 4V) applied to the tungsten tip of the piezo-controlled nanomanipulator, which serves both as a mechanical manipulator and mobile electrode[18]. As the tungsten tip moves to the left (Fig. 1(b)) it extracts one or more walls of the parent tube, yielding the daughter tube which telescopes or slides off of the fixed core. The goal is to produce a daughter tube that is susceptible to collapse, i.e. one that has few walls and a relatively large outer diameter (limited, of course, by the outer diameter of the parent tube). When a suitable length of daughter tube has been extracted (Fig. 1(c)), collapse of the daughter tube is mechanically initiated by a small displacement of the tungsten tip transverse to the tube axis (vertical arrow in Fig. 1(c)). This typically leads to a runaway zipper-like collapse process, where the daughter tube first collapses locally at the perturbation site, and the collapsed region then quickly propagates the length of the daughter tube, where collapse is halted by the core plug. The overall length of the collapsed daughter tube can be further and controllably extended by additional movement to the left of the tungsten tip (as the daughter tube telescopes from the stationary core, the newly exposed "hollow" region immediately collapses via the zipper effect still driven by van der Waals forces). The maximum length of the doubly-clamped collapsed daughter tube is equal to the total length of the parent multi-wall tube (Fig. 1(e)). Further movement of the tungsten tip to the left yields full detachment on the right of the daughter tube from the core, resulting in a CCNT singly clamped at one end and suspended in vacuum at the other end. Spontaneous twisting can then occur unimpeded (Fig. 1(f)). We note that it is possible to reattach, via *in-situ* spot welding, the right end of the daughter CCNT to the remaining core tube or the copper mesh TEM grid itself, resulting in a doubly-clamped twisted CCNT (Fig. 1(g)). Such reattachment of the free end aids in TEM imaging of the twisted CCNT by largely quenching vibrations.

The right panel of TEM images in Fig. 1 illustrates an actual experiment performed as outlined above (employing a JEOL 2010 TEM operated at 80 keV), with direct correspondences between the TEM images and the adjacent schematic drawings. The inflated parent tube (visible on the right side of Fig. 1(h)) has about 31 walls and is 26 nm in outer diameter. The tailored daughter tube has three walls and an outer diameter of 26 nm; such a daughter tube is very susceptible to collapse as it is more than three times wider than the critical radius of a three-walled tube (7.6 nm).[6] The arrow in Fig. 1(i) indicates where collapse of this tube was initiated via mechanical perturbation.  (We note that the extreme left end of the daughter nanotube is prevented from completely collapsing due to the tube end being held open there by the spot weld to the tungsten tip.)

The width of the daughter nanotube after collapse is 39 nm, as determined from the TEM image of Fig. 1(j)), consistent with what is expected for a collapsed state.[19] When the CCNT is nearly completely detached from the inner core, we observe narrowing at the center (Fig. 1(k)) due to folding of edges towards the tube axis. When the tube is fully detached from the inner core and is in a singly-clamped condition, it twists along its axis with two nodes. Fig. 1($\ell$) shows the twisted condition clearly (we note that for Fig. 1($\ell$) the right end of the nanotube has been reattached to the stationary core on the right to minimize the vibration for enhanced TEM characterization).

The ability to characterize structurally a given nanotube before and after collapse is essential for elucidating why CCNTs twist.  We analyze the lattice registry of the tailored CCNT of Fig. 1 before and after twisting by selected area electron diffraction (SAED). The SAED pattern (Fig. 2(b)) taken at the right end of the CCNT - where it is partially collapsed and partially tubular (outlined by a blue circle in Fig. 2(a)) - shows only two sets of six-fold graphitic diffraction spots corresponding to the two opposing graphitic walls of the nanotube. This indicates that all three walls of the daughter nanotube have the same chirality. The streaked features in the diffraction spots are due to the curvature of the inflated part of the tube, while more defined round diffraction spots come from the flat region of the collapsed part. Since the graphitic diffraction spots for the inflated and collapsed parts are nearly identical, we conclude that the opposing faces of the inner walls maintain their orientation during collapse.   The numerical value of the chiral angle for the inflated daughter tube is $\theta = 27.1\pm0.5°$ [20]  which is consistent with the relative rotation angle between graphitic layers of the collapsed part ($\beta=7\pm0.5°$).  Even more importantly, we find that the orientation is nearly unchanged even after twisting.  This can be seen from a SAED pattern (Fig. 2(d)) obtained from the center of the twisted CCNT (Fig. 2 (c)) which gives nearly the same orientation (β ~6°) as that before twisting (β ~7°).

We now examine *registry*[12,14,15] and *edge*[16] mechanisms for the twisting of a CCNT. In the registry mechanism twisting in a CCNT is driven by a lattice registry effect between opposing walls. Attainment of the most stable energy configuration drives local lateral shifting of the opposing walls of the CCNT, which then manifests itself as twisting along the axis of the CCNT. The edge mechanism for twisting is adapted from mechanical stability studies of a freestanding single layer graphene nanoribbon[16], or, analogously, from instabilities well known for macroscopic sheet metal strips processed via rollers[21,22]. Here twisting is attributed to a non-uniform stress in the ribbon resulting from elongation of the ribbon edge material relative to the center material. In graphene nanoribbons the length differential originates from lattice reconstruction of the edge atoms[16,17], while for sheet metal it results from non-planar rollers that non-uniformly elongate the material[21]. As we demonstrate below neither the registry nor the edge mechanism accounts for our observations, and we present an alternate model.

Our theoretical examination of competing mechanisms is derived from a continuum elasticity model that we justify by a first-principles density functional theory (DFT) calculation, using the quantum espresso package [23]. First, we simplify our calculation by showing that the bulged edges of a typical CCNT have very little effect on its overall elastic behavior. We justify this simplification by two DFT calculations of elastic properties relevant for twisting: First we calculate the energy cost per atom of a uniaxial compression for pristine graphene and for a CCNT, and find them to agree better than 1% in a range of strain from -1.5% to +1.5%. A second quantity we check is the softening of the out-of-plane (flexural) phonon mode by a uniaxial compression. As found in Ref.[24] this phonon softening mechanism causes buckling of a pristine graphene sheet. We find that softening of a flexural phonon is nearly the same (by less than 20%) in graphene as in a CCNT. The discrepancy here is larger than in the first case likely because we approximate the flexural phonon eigenvector in a CCNT in order to render the calculation manageable[25].

Therefore, in the following analysis of the CCNT twisting we will ignore the effects of the bulges and treat the CCNTs as a stack of flat pristine graphene sheets. Elastic properties of a pristine graphene sheet (and therefore of a CCNT) are well described with a continuum model[26]. To lowest order of the elastic continuum model we obtain the following areal energy density ($E$) of a twisted CCNT with width $W$ and twist period $T$,

$$E = \frac{\pi^2}{6} \epsilon\, C \left(\frac{W}{T}\right)^2 + \frac{\pi^4}{40} C \left(\frac{W}{T}\right)^4. \qquad (1)$$

The only material dependent quantity here is the elastic modulus $C$, which we find equal to 23 eV/A$^2$ both in graphene and in CCNTs (within 0.2%). The parameter $\epsilon$ in Eq. (1) represents an effective strain along the long nanotube axis either due to stress or charge doping (effects of stress and doping are indistinguishable in the continuum model).

We use Eq. (1) to calculate the elastic energy cost of twisting. Inserting the parameters for the tailored daughter nanotube $W$ (39 nm) and $T$ (600 nm) into the second term (and setting $\epsilon$ equal to zero) shows that twisting of the CCNT costs only 2.8 meV per each carbon atom. In the registry mechanism proposed in Refs. [14,15] this increase in the elastic energy is compensated by a change in the lattice registry between the opposing graphitic walls of the CCNT. However, as discussed earlier, our *in-situ* TEM study shows nearly no change in the relative rotation angle between the opposing walls upon twisting (it is 7° before and 6° after twisting). There is no substantial energy difference between those two lattice mismatch angles since in both cases lattice registry is an equal admixture of AA and AB stacking. Even if we consider for a moment an extreme case where interlayer interaction changes from β = 7° all the way to the AB stacking (β = 0°) we estimate that the energy reduction is at most only 0.8 meV per atom, more than three times smaller than the change in the twisting elastic energy. Therefore we conclude that lattice registry is not responsible for twisting of a CCNT. (This estimate takes into account that out of the three CNT walls, only the inner-most interface has a change in the lattice registry and that the energy difference between AA and AB stacked graphite is 10 meV per carbon atom.)

We now turn to the edge mechanism, i.e. the possibility that the edges of the CCNT behave differently than the bulk resulting in, for example, a differential elongation and consequent rippling and/or twisting. The CCNT in question is quite large (39 nm wide) with only about 5% of the carbon atoms are at the edge of the tube, hence the energy reduction at the edge would have to be quite large (56 meV per atom) to compensate the energy increase of the twist. Therefore, although this mechanism may be appropriate in describing rippling/twisting in narrow graphene *nanoribbons*, it is ruled out for twisting in our typical CCNTs.

We now discuss specifically our proposed mechanism for CCNT twisting. An important feature of Eq. (1) is that the restoring energy cost of the twisting is proportional to the fourth power of *W/T* while the effect of the strain $\epsilon$ due to external loading is proportional to the second power of *W/T*. Therefore even a slightest negative $\epsilon$ will twist a CCNT. In other words, total energy *E* is minimized for a finite twist period T satisfying relation

$$\frac{W}{T} = \frac{\sqrt{10}}{\pi\sqrt{3}}\sqrt{-\epsilon}. \qquad (2)$$

This relation is graphically represented a purple line in in Fig. 3. The insets in Fig. 3 show the energy density (E) of a CCNT vs. the W/T ratio both under compressive (left graph) and tensile (right graph) strain.

Inserting *W* (39 nm) and *T* (600 nm) into Eq. (2) we find that an effective compressive strain of only 1.3% (either due to stress or doping) is needed to produce our observed twisting of the CCNT. In a typical collapsing/twisting scenario occurring in the nanotube synthesis chamber, non-local external stresses from extreme temperature gradients, collisions with high energy ions, gas flow, etc. are likely a common occurrence and account for "naturally occurring" twisted CCNTs[1]. In our tightly controlled clamped-free CCNT configuration of Fig. 1, a strain $\epsilon$ occurs most likely from the compression caused by a charge imbalance. Charge imbalance can result from intrinsic doping (common to multi-wall carbon nanotubes), with some contribution also possible from the TEM imaging beam itself.

Interestingly, while the continuum model predicts that a CCNT will twist even under an infinitesimal external load (see Fig. 3), previous work from Ref.[24] predicts that a related distortion, rippling of a graphene sheet, occurs only when load is above some critical value.

**Acknowledgements:**

This work was supported in part by the Director, Office of Basic Energy Sciences, Materials Sciences and Engineering Division, of the U.S. Department of Energy under Contract #DE-AC02-05CH11231, within the Nanomachines Program, which provided for TEM characterization and for the continuum model calculation; by the Office of Naval Research under contract N00014-12-1-1008 which provided for collapsed nanoribbon synthesis; by the National Science Foundation under grant DMR15-1508412 which provided for total energy calculations, and by the Swedish Research Council (grant dnr 2015-00520) which provided support for HRB. Computational resources have been provided by the NSF through XSEDE resources at NICS.

**Figures:**

**Figure 1:** Schematic of our experimental setup along with corresponding TEM images on the right. As the W tip is moved to the left it strips off a large diameter (26 nm) three-walled CNT that first collapses and then twists. For graphical convenience, images have been displaced laterally so that the (mobile) W tip always remains at the left edge of the image. The dashed arrows which point to the left in (b) (c) and (d) indicate the direction the W tip actually moves. The other dashed arrow which points vertically in (c) indicates the direction of the mechanical perturbation from the W tip. The solid arrow in (i) points to the kink formed by the mechanical deformation from the W tip. Black scale bar for panels (h)-(l) is 100 nm.

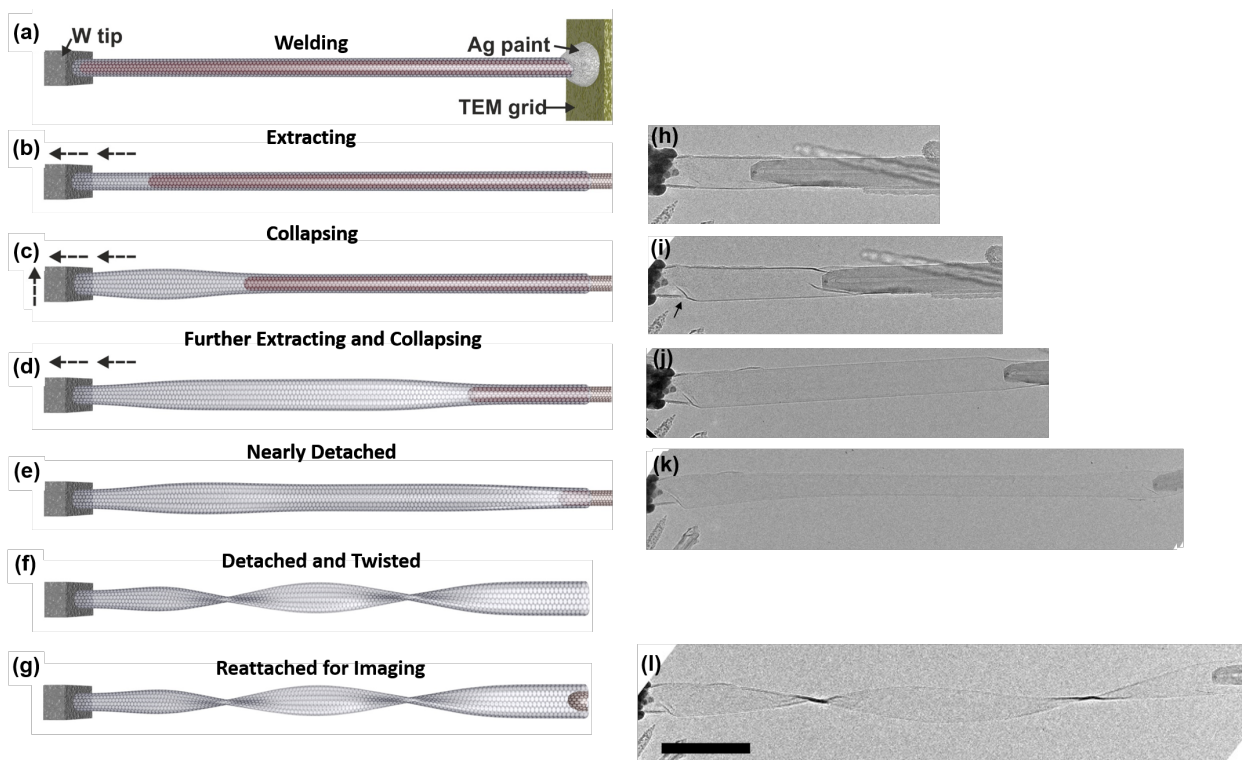

**Figure 2:** TEM image of the collapsed CNT before twisting (a) and SAED pattern (b) of a partially collapsed region outlined by a blue circle (left side of the circle is collapsed and right side is not). TEM image after twisting (c) and SAED pattern (d) taken from the central part of the tube. Scale bars in (a) and (c) are 100 nm.

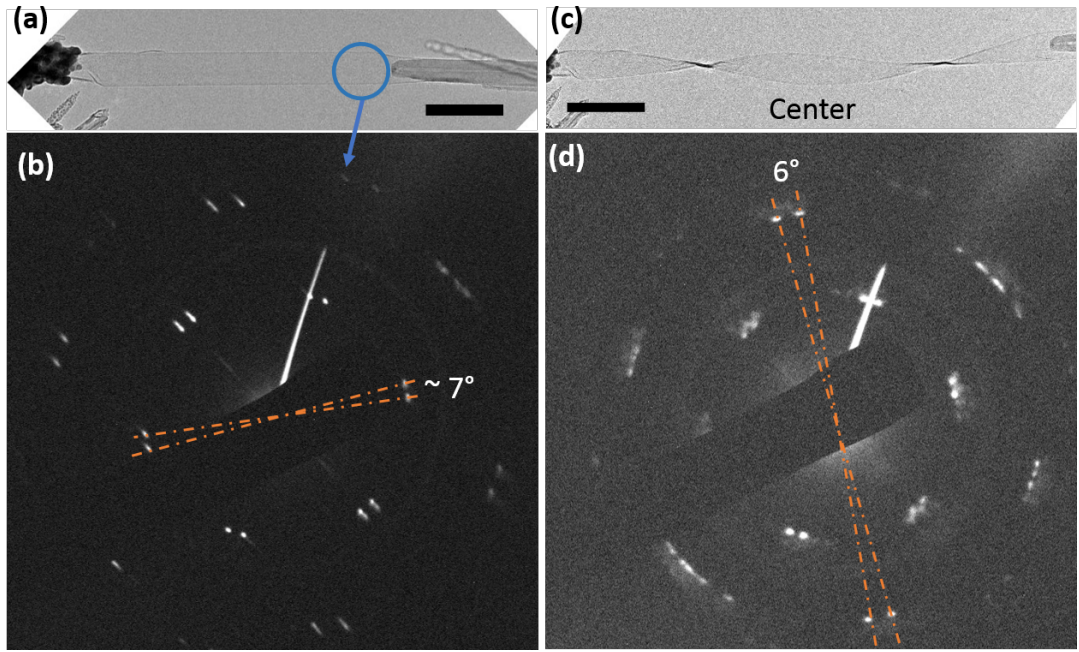

**Figure 3:** Our continuum model predicts that the CCNT will twist under an infinitesimal compressive strain $\epsilon < 0$. Two small inset graphs show dependence of energy (E) on the ratio of CCNT width (W) and the twist period (T) both under compressive (red graph) and tensile strain (blue graph). The purple line shows the ratio W/T from Eq. (2) that minimizes the energy of the CCNT.

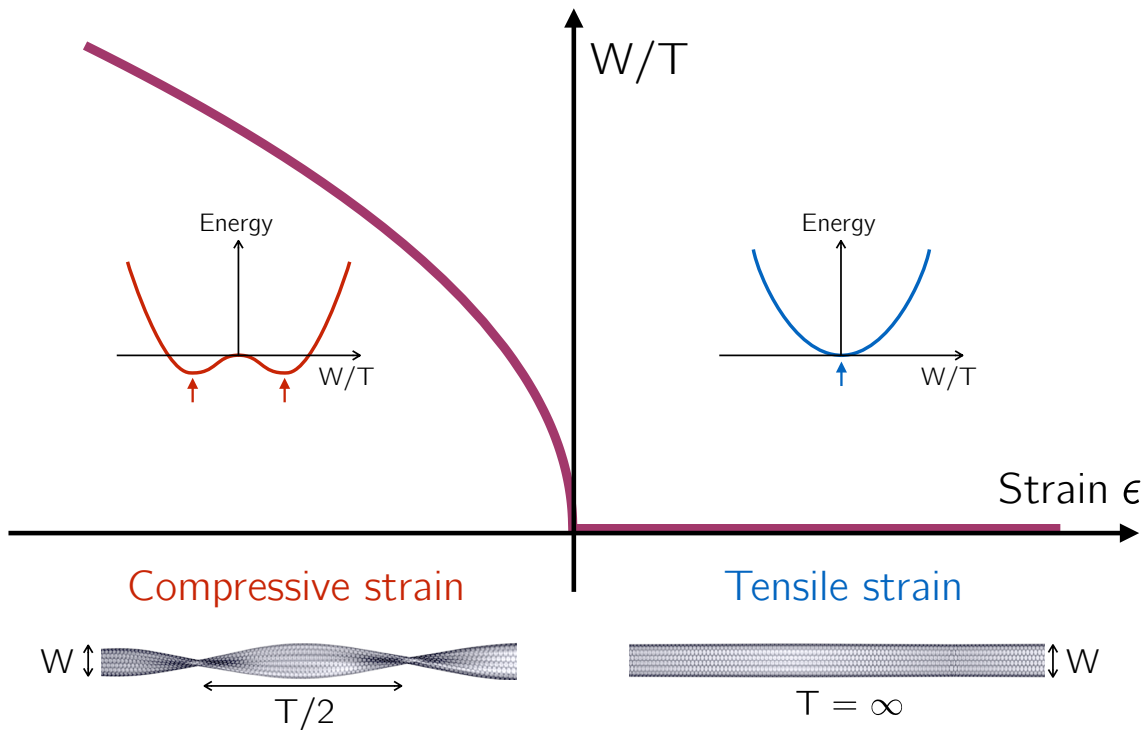